\documentclass[%
reprint,
superscriptaddress,
showpacs,preprintnumbers,
 amsmath,amssymb,
 aps,
prd,
]{revtex4-1}

\usepackage{graphicx}
\usepackage{dcolumn}
\usepackage{bm}
\usepackage{bbold}
\usepackage{amssymb,amsmath}
\usepackage{hyperref}

\usepackage{color}
\usepackage{amsfonts}
\usepackage{subfigure}
\usepackage{array}

\newcommand{\tr}{\ensuremath{\operatorname{tr}}}

\newcolumntype{L}{>{\centering\arraybackslash}m{3cm}}

\definecolor{bjcol}{rgb}{1,.44,0.13}

\definecolor{blue}{rgb}{0,0,1}

\definecolor{green}{rgb}{0,1,0}

\definecolor{red}{rgb}{1,0,0}

\definecolor{gray}{rgb}{.5,.5,.5}

\definecolor{darkgreen}{rgb}{.0,.5,.0}

\def\Fig#1{Fig.~\ref{#1}} \def\Tab#1{Tab.~\ref{#1}}
 \def\Tab#1{Tab.~\ref{#1}}

\def\Eq#1{Eq.~(\ref{#1})}
\def\eq#1{(\ref{#1})}
\def\eqref#1{(\ref{#1})}

\def\tab#1{Tab.~\ref{#1}}

\def\lA0{{\langle A_0 \rangle}}
\def\bA0{{\bar{A}_0}}

\def\pslash{p\llap{/}}

\def\0#1#2{\frac{#1}{#2}}

\graphicspath{{./figures/}{./}}

\begin{document}

\preprint{}

\title{Baryon number fluctuations at finite temperature and density
}

\author{Wei-jie Fu}
\affiliation{Institute of Theoretical Physics, School of Physics \&
  Optoelectronic Technology, Dalian University of Technology, Dalian, 116024,
  P.R. China}
\affiliation{Institut f\"{u}r Theoretische Physik, Universit\"{a}t
  Heidelberg, Philosophenweg 16, 69120 Heidelberg, Germany}

\author{Jan M. Pawlowski}
\affiliation{Institut
 f\"{u}r Theoretische Physik, Universit\"{a}t Heidelberg,
 Philosophenweg 16, 69120 Heidelberg, Germany}
\affiliation{ExtreMe Matter Institute EMMI, GSI, Planckstr. 1, 64291
  Darmstadt, Germany}

\author{Fabian Rennecke}
\affiliation{Institut
 f\"{u}r Theoretische Physik, Universit\"{a}t Heidelberg,
 Philosophenweg 16, 69120 Heidelberg, Germany}
\affiliation{Institut f\"{u}r Theoretische Physik, Justus-Liebig-Universit\"at Gie{\ss}en,
Heinrich-Buff-Ring 16, 35392 Gie{\ss}en, Germany}

\author{Bernd-Jochen Schaefer}
\affiliation{Institut f\"{u}r Theoretische Physik, Justus-Liebig-Universit\"at Gie{\ss}en,
Heinrich-Buff-Ring 16, 35392 Gie{\ss}en, Germany}


\begin{abstract}

  We investigate baryon number fluctuations for finite temperature and
  density in two-flavor QCD. This is done within a QCD-improved
  low-energy effective theory in an
  extension of the approach put forward in
  \cite{Fu:2015naa,Fu:2015amv}. In the present work we aim at
  improving the predictive power of this approach for large
  temperatures and density, that is, for small collision energies. This
  is achieved by taking into account the full frequency dependence of
  the quark dispersion. This ensures the necessary Silver Blaze
  property of finite density QCD for the first time, which so far was
  only implemented approximately. Moreover, we show that Polyakov loop
  fluctuations have a sizeable impact at large temperatures and
  density. The results for the kurtosis of baryon number fluctuations
  are compared to previous effective theory results, lattice results
  and recent experimental data from STAR.

\end{abstract}

\maketitle


\section{\label{sec:intr}Introduction}

The past years have seen rapid progress of our understanding of
heavy-ion collision physics and QCD in extreme conditions. This
progress has been achieved by both experimental measurements at the
Relativistic Heavy-Ion Collider (RHIC) and the Large Hadron Collider
(LHC), as well as theoretical calculations made in various ab initio
and effective theory approaches. One of the remaining key challenges
is to get hold of the existence and location of the critical end point
(CEP) in the QCD phase diagram
\cite{Stephanov:2007fk}. Experimentally, the search of the CEP is
under way in the Beam Energy Scan (BES) program at RHIC
\cite{Adamczyk:2013dal,Adamczyk:2014fia,Luo:2015ewa}, as well as
future searches at the FAIR and NICA facilities, and the evaluation of
current results at low collision energies at HADES, spanning a wide
collision energy or density regime, see
\cite{Chattopadhyay:2016qhg}. On the theoretical side, first principle
lattice computations are hampered by the notorious sign problem at
finite chemical potential \cite{Aarts:2015tyj}. First principle
functional continuum computations are hampered by the task of
systematically taking into account the relevant degrees of freedom
\cite{Pawlowski:2014aha}. This calls for refined effective theory
investigations that are embedded in QCD such, that they allow for a
systematic improvement towards full QCD. This approach is taken in the
current work, extending the recent works \cite{Fu:2015naa,Fu:2015amv}.

Correlations of conserved charges provides good experimental
signatures of the CEP: as the transition at the CEP is of second
order, a singularity is expected in thermodynamic quantities. However,
since the QGP produced in heavy-ion collisions is finite both
spatially and temporally, such singularities cannot be observed in
Nature. Nonetheless, fluctuations in event-by-event multiplicity
distributions of conserved quantities such as variances and moments of
these distributions become more sensitive around the CEP since their
criticality are proportional to powers of the correlation length
\cite{Stephanov:1998dy, Stephanov:1999zu, Stephanov:2008qz}. This
intuitive physical picture is the foundation for present and future
experimental searches for the critical end point. In the present work,
we investigate baryon number fluctuations which are described by the
generalised susceptibilities and are given by derivatives of the
pressure $p$ with respect to the baryon chemical potential $\mu_B = 3
\mu$.  Thus, the accurate description of baryon number fluctuations
requires a thorough understanding of the chemical potential dependence
of the equation of state of QCD. But this is still a formidable
problem in theoretical QCD investigations and far from being solved.

Furthermore, since the created quark-gluon plasma is of finite spatial
extent and cools down rapidly, the system evolves out of equilibrium
in the vicinity of the CEP. Due to the critical slowing down
phenomenon long equilibration times are expected and the correlation
length cannot grow as fast as its equilibrium counterpart in the
expanding plasma \cite{Berdnikov:1999ph}. As a consequence, the
generalised susceptibilities measured in the experiment could differ
in both magnitude and sign from the equilibrium prediction in the
critical region \cite{Mukherjee:2015swa,Mukherjee:2016kyu}.

Nonetheless, understanding the chemical potential dependence of the
equation of state in equilibrium is a crucial building block for a
deeper comprehension of the signatures of the CEP in heavy-ion
collisions. Recently, some of us have investigated the QCD
thermodynamics, the skewness and the kurtosis of the baryon number
distributions within QCD-improved low-energy effective models
\cite{Fu:2015naa,Fu:2015amv}, for related work see also
\cite{Borsanyi:2013hza,Ding:2014kva, Skokov:2010wb, Skokov:2010sf,%
Skokov:2010uh, Karsch:2010hm, Fu:2009wy, Fu:2010ay, Schaefer:2011ex,%
Schaefer:2012gy,Morita:2014fda, Morita:2014nra}. In these
computations quantum, thermal, and density fluctuations are embedded
with the functional renormalisation group (FRG) approach to QCD, see
e.g. \cite{Haas:2013qwp, Herbst:2013ufa,Pawlowski:2014zaa,%
Helmboldt:2014iya, Mitter:2014wpa, Braun:2014ata,Pawlowski:2014aha}
and references therein. In the present work, we significantly improve 
on these previous studies.

The chemical potential dependence of the equation of state of QCD is
intimately linked to a peculiar feature of finite density QCD known as
the Silver Blaze property \cite{Cohen:2003kd}. It states that at
vanishing temperature observables are independent of the chemical
potential below a critical one. A proper description of QCD at finite
chemical potential has to respect this property. In the context of the
functional renormalisation group, it was shown that the Silver Blaze
property is directly linked to the frequency dependence of correlation
functions that involve particles with non-vanishing baryon number
\cite{Khan:2015puu,Fu:2015naa}. We have generalised the previous works
and have implemented for the first time frequency dependent quark
correlation functions to the equation of state. This fully guarantees
the Silver Blaze property in such a fluctuation analysis. As we shall
see these modifications have a significant effect on both the
magnitude and the sign of the kurtosis of baryon number fluctuations
at finite density.

Another related crucial issue in this respect is how the gluon
fluctuations and confinement affect the baryon number fluctuations. In
the low-energy sector of QCD, gluon effects are implemented in a
non-vanishing gluon background field whose thermodynamics is encoded
in a Polyakov loop potential. It has been argued in \cite{Fu:2015naa}
that the baryon number susceptibilities are rather sensitive to
Polyakov loop fluctuations. Hence, we include a phenomenological
Polyakov loop potential that captures the effect of Polyakov loop
fluctuations \cite{Lo:2013hla}. Such a potential incorporates the
back-reaction of the gluon effects on the matter sector of QCD.  This
significantly influences the baryon number fluctuations at
temperatures above $T_c$.

The paper is organised as follows: In Sec.~\ref{sec:FRG} we briefly
introduce the approach of QCD-improved low-energy effective theories
within the FRG framework. In Sec.~\ref{sec:fre} the flow equations in
the presence of a frequency dependent quark anomalous dimension are
discussed and some implications are discussed. Numerical results,
their discussion and comparison with lattice and experimental data are
provided in Sec.~\ref{sec:num}. A summary with our conclusions can be
found in Sec.~\ref{sec:sum} and technical details on the used
threshold functions are collected in the appendix.


\section{\label{sec:FRG} FRG for QCD and the low energy effective theory}

In this work we improve the previous studies
\cite{Fu:2015naa,Fu:2015amv} on baryon number fluctuations within a
low-energy effective theory in two aspects: Firstly, we extend the
approximation to the off-shell fluctuation physics used in the
previous works.  This is important for both quantitative precision as
well as the systematic error control in the present approach. This
leads us to an improved effective potential, where the
fluctuation-induced and frequency-dependent corrections to the quark
dispersion are taken into account.  These improvements are important
for the proper description of baryon number fluctuations in particular
at finite chemical potential due to the intimate relation between the
frequency dependence of correlation functions and the chemical
potential dependence of the theory.  This will be elaborated in detail
in the following section. Secondly, the improved fluctuation analysis
is extended to the glue sector. Here we incorporate a Polyakov loop
potential which captures the Polyakov loop
fluctuations \cite{Lo:2013hla}.  Such a potential takes into
  account the impact of off-shell fluctuations beyond the level of
  expectation values of the Polyakov loop variable and the
  thermodynamics. Such an extension has been argued to be important in
  \cite{Fu:2015naa} for the evaluation of the baryon number
fluctuations.

We begin with the discussion of the effective model. It is based on
the description of low-energy QCD for two flavors put forward in
\cite{Herbst:2010rf,Pawlowski:2014zaa, Fu:2015naa}. Here, we give a
brief summary and refer the interested reader to the literature for
details. In order to capture the relevant hadronic degrees of freedom
at small and intermediate densities, we include the pion and the sigma
mesons as the dominant low-energy degrees of freedom. They are coupled
to quarks via a Yukawa interaction term with a running coupling
$h_k$. The purely mesonic interactions are stored in the effective
potential $V_k(\rho)$ with $\rho = (\vec{\pi}^2 + \sigma^2)/2$.
Fluctuation-induced corrections to the classical quark and meson
dispersion relations are taken into account by the corresponding quark
and meson wave function renormalisations $Z_{q,k}$ and
$Z_{\phi,k}$. Due to the dynamical generation of the gluon mass gap,
the gluon sector of QCD decouples at low energies
$ \Lambda \lesssim 1$ GeV and the informations of the deconfinement
transition are encoded in a non-vanishing gluon background field for
energies $k<\Lambda$. We therefore introduce a non-vanishing temporal
gluon background field $A_0$ which couples to the quarks as well as to
a corresponding effective potential $V_\text{glue}(L,\bar L)$. The
potential is formulated in terms of the expectation value of the
traced Polyakov loop $L$ and its adjoint $\bar L$.  They are given by
\begin{align}\label{eq:Lloop}
  L(\vec{x})=\0{1}{N_c} \left\langle \tr\, {\cal P}(\vec
    x)\right\rangle \,,\quad \quad \bar L (\vec{x})=\0{1}{N_c} \langle
  \tr\,{\cal P}^{\dagger}(\vec x)\rangle \,,
\end{align}
with 
\begin{align}\label{eq:Ploop}
  {\cal P}(\vec x)= \mathcal{P}\exp\Big(ig\int_0^{\beta}d\tau
    A_0(\vec{x},\tau)\Big)\,.
\end{align}
We postpone the introduction and discussion of the Polyakov loop
potential $V_\text{glue}(L,\bar L)$ to the next section.

Such a construction results in a Polyakov--quark-meson (PQM) model
\cite{Schaefer:2007pw} and the corresponding effective action reads
\begin{align}\nonumber 
  &\Gamma_{k}=\int_{x} \Big\{Z_{q,k}\bar{q}
  \big[\gamma_{\mu}\partial_{\mu}
  -\gamma_{0}(\mu+igA_0)\big] q\, + V_\text{glue}(L,\bar L)\\[2ex]
  &+\frac{1}{2}Z_{\phi,k}(\partial_{\mu}\phi)^2 +h_{k} \,\bar{q}\left(
    T^{0}\sigma+i\gamma_{5} \vec{T}\vec{\pi}\right)
  q+V_{k}(\rho)-c\sigma\Big\}\,,
\label{eq:action}\end{align}
with $\int_{x}=\int_0^{1/T}d x_0 \int d^3 x$ and the quark chemical
potential $\mu$. The gluonic background field is constant and only its
temporal component assumes a non-vanishing expectation value. The real
meson field $\phi=(\sigma,\vec{\pi})$ is in the $O(4)$-representation,
and $\rho= \phi^2/2$.  The $SU(N_{f})$ generators $\vec{T}$ are
normalised as $\mathrm{tr}(T^{i}T^{j})=\frac{1}{2}\delta^{ij}$ with
$T^{0}=\frac{1}{\sqrt{2N_{f}}}\mathbb{1}_{N_{f}\times N_{f}}$. The
chiral effective potential $V_{k}(\rho)$ is $O(4)$ invariant, and the
linear term $-c\sigma$ breaks chiral symmetry explicitly.  Note that
the matter and the gauge sector of QCD are naturally coupled to each
other by considering a non-vanishing gluon background.  The background
field $A_0$ enters the chiral effective potential $V_k$ through quark
fluctuations \cite{Fukushima:2003fw}. In this way the correct
temperature scaling of the gluon potential in QCD is recovered: at
vanishing density and finite current quark masses, both the chiral and
the deconfinement transition are smooth crossovers.  This also holds
for thermodynamical quantities like the pressure and trace
anomaly. With such a QCD-enhanced glue potential and for $N_f=2+1$
quark flavor nice agreement with recent lattice QCD results can be
obtained, see \cite{Herbst:2013ufa}.

\begin{figure}[t]
\includegraphics[width=0.4\textwidth]{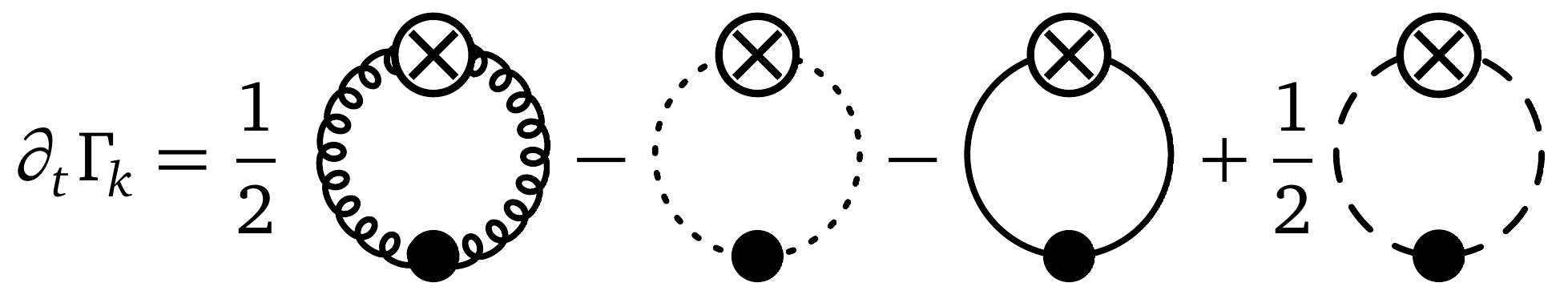}
\caption{Flow of the effective action. The first two diagrams are the
  gluon and ghost contributions. Here, they are assumed to be
  integrated out. The low-energy information of the gluon sector is
  stored in an effective Polyakov loop potential. The last two
  diagrams are the matter contributions to the flow, i.e. quarks and
  mesons in the present case. The black dots indicate that the fully
  dressed propagators are involved in the flow equation. The crossed
  circles denote the regulator insertion.  }\label{fig:fleq}
\end{figure}

All couplings in the effective action depend on the renormalisation
group (RG) scale $k$. By following the evolution of $\Gamma_k$ from
the UV cutoff scale $k=\Lambda$ down to the infrared $k=0$, quantum
fluctuations are successively included in the effective action. The
evolution equation for $\Gamma_k[\Phi]$, where
$\Phi=(A_\mu,c,\bar c,q,\bar q,\phi,...)$ indicates the super field,
is given by the Wetterich equation \cite{Wetterich:1992yh},
\begin{align}
  \label{eq:dtGam}
  \partial_{t}\Gamma_{k}[\Phi]=\frac{1}{2}\mathrm{Tr}\,G_{\Phi
    \Phi}[\Phi]\partial_{t} R^{\Phi}_{k}\,, \quad t=\ln (k/\Lambda)\,,
\end{align}
with the exact field-dependent propagator 
\begin{align}\label{eq:GPhi} 
  G_{\Phi_i \Phi_j}[\Phi] =
  \left(\0{1}{\0{\delta^2\Gamma_k[\Phi]}{\delta\Phi^2}+R_k^\Phi}\right)_{ij}\,,
\end{align}
and a regulator $R_k^{\Phi}$.  Within this framework, it is by now
well understood how the used effective low-energy model is embedded
into full QCD. To that end we rewrite the effective action as
\begin{align}\label{eq:Gasplit}
  \Gamma_k[\Phi]=
  \Gamma_{\text{\tiny{glue}},k}[\Phi]+\Gamma_{\text{\tiny{matt}},k}[\Phi]\,,
  \quad
  \Gamma_{\text{\tiny{matt}},k}=\Gamma_{q,k}+
\Gamma_{\phi,k}\,,
\end{align}
where $\Gamma_{\text{\tiny{glue}},k}$ encodes the ghost- and gluon
fluctuations and is the glue sector of the effective action. The
matter sector $\Gamma_{\text{\tiny{matt}},k}$ is composed of
$\Gamma_{q,k}[\Phi]$ arising from quark fluctuations, and
$\Gamma_{\phi,k}[\Phi]$ from that of the hadronic degrees of freedom,
see Fig~\ref{fig:fleq}. The separation between the quark and hadronic
contributions is realised through the dynamical hadronisation
\cite{Gies:2001nw,Gies:2002hq,Pawlowski:2005xe,Floerchinger:2009uf}, a
very efficient parameterisation of matter fluctuations in ab initio
QCD, for applications to QCD see e.g.\
\cite{Mitter:2014wpa,Braun:2014ata}.

%
%

The gluon and ghost fluctuations start to decouple from the matter
sector after the QCD-flow is integrated down to scales $k=\Lambda$
with $\Lambda\lesssim 1$ GeV, for more details, see e.g.,
\cite{Mitter:2014wpa,Braun:2014ata}. We are left with an effective
matter theory with a glue background, which is given by
Polyakov-loop--extended chiral models, such as the
Polyakov--Nambu--Jona-Lasinio
model~\cite{Fukushima:2003fw,Ratti:2005jh,Fu:2007xc} and the
Polyakov--quark-meson (PQM) model~\cite{Schaefer:2007pw}. For further
details in this direction, we refer to recent works and reviews,
e.g. \cite{Pawlowski:2014aha, Mitter:2014wpa, Braun:2014ata,
  Rennecke:2015eba,Cyrol:2016tym}.

In summary, below the decoupling scale of the glue sector we are left
with an effective theory that is well described with an effective
action \eq{eq:action}. Further quantum, thermal and density
fluctuations are then obtained by the flow equation with the remaining
dynamical degrees of freedom, the quarks and mesons. Within the
present approximation this boils down to the low-energy QCD-flows
used in \cite{Braun:2014ata} in the vacuum. For the computation of
baryon number fluctuations derivatives of the pressure with respect to
the quark chemical potential are needed. The pressure is extracted
from the scale dependent effective action $\Gamma_k$, \eq{eq:action},
in the infrared via the thermodynamic potential
\begin{align}\label{eq:Omega}
\Omega[\Phi;T,\mu] = V_\text{glue}(L,\bar L) + V_{k=0}(\rho)-c\sigma\,. 
\end{align}
\Eq{eq:Omega} simply constitutes the effective action $\Gamma_{k=0}$ at
vanishing cutoff scale $k$, evaluated on constant backgrounds $L,\bar
L,\sigma$.  Finally the backgrounds are chosen such that they solve
the equations of motion (EoM): $\Phi_{\text{\tiny{EoM}}}$. The rest
of the fields vanish on the EoMs, so they are taken to be zero
straightaway. Then the normalised pressure reads
\begin{align}
  p(T,\mu) = -\Omega[\Phi_{\text{ \tiny{EoM}}};T,\mu] +
  \Omega[\Phi_{\text{\tiny{EoM}}};0,0]\,.
\end{align}
Hence, by solving the flow equation \eq{eq:dtGam} for the effective
action \eq{eq:action}, we extract the pressure and obtain the baryon
number fluctuations from appropriate $\mu$-derivatives. The latter are
defined by the generalised susceptibilities
\begin{equation}
\label{eq:bnf}
\chi_n^{\mathrm{B}}=\frac{\partial^n}{\partial (\mu_{\mathrm{B}}/T)^n}\frac{p}{T^4}\,,
\end{equation}
which are given as $n^{\text{th}}$-derivatives of the pressure $p$
with respect to the baryon chemical potential $\mu_{\mathrm{B}}$
related to the quark chemical potential by $\mu=\mu_{\mathrm{B}}/3$.

Cumulants of baryon multiplicity distributions, which can be measured
experimentally, are closely related to the generalised
susceptibilities $\chi_n^{\mathrm{B}}$, such as the variance
$\sigma^2=VT^3\chi_2^{\mathrm{B}}$ or the kurtosis
$\kappa=\chi_4^{\mathrm{B}}/(\chi_2^{\mathrm{B}}\sigma^2)$. All
generalised susceptibilities depend on the volume $V$ of the system
which drops out by considering ratios like the kurtosis of the
susceptibilities.


\subsection{Fluctuations and the Polyakov Loop Potential}
\label{sec:polloop}

It follows from the discussion of the last chapter that the glue part
of the potential cannot be obtained within the present effective
theory approach, which only considers quark and meson fluctuations
below the glue decoupling scale. Moreover, the quarks couples to the
mean temporal gauge field, $\langle A_0\rangle$, rather than the mean
Polyakov loop $L$ defined in \eq{eq:Lloop}. For a related detailed
discussion and a computation and comparison of both observables see
\cite{Herbst:2015ona}. 

In the present work we shall ignore this issue, and instead resort to
utilising pure glue lattice data. In absence of a lattice computations
of the Polyakov loop potential $V(L,\bar L)$ in $SU(3)$, Yang-Mills
lattice data on correlation functions are used. This includes the
thermal pressure $p$, the Polyakov loop expectation value $L$, and the
fluctuations, e.g.\ $ \langle\tr \mathcal{P}(\vec{x}) \tr
\mathcal{P}(\vec{y}) \rangle$ with $\cal P$ defined in \eq{eq:Ploop}. These
observables determine the minimum of the potential ($L$), the value of
the potential at the minimum ($p)$, as well as all second derivatives
$\partial^2_{L,\bar L} V$ w.r.t.\ $L$, $\bar L$ at the
minimum. Finally, the temperature scales in the pure glue potentials
have to be adapted to full dynamical QCD as has been put forward in
\cite{Haas:2013qwp, Herbst:2013ufa}. In the present context this has
been discussed in \cite{Fu:2015naa}. In summary it amounts to the
rescaling of the reduced temperature $t=T/T_0$ in the Yang-Mills
potential by a factor $0.57$. Here $T_0$ is the Yang-Mills critical
temperature. In the present set-up the absolute temperature scale
$T_0$ is fixed by the requirement of equivalent
confinement-deconfinement pseudo-critical and chiral critical
temperatures in the chiral limit as predicted in \cite{Braun:2009gm}.

It is left to choose the specific parameterisation of the pure glue
potential.  There are various possibilities to model such a Polyakov
loop potential. Most commonly used potentials are either of polynomial
\cite{Ratti:2005jh} or logarithmic form \cite{Fukushima:2008wg}. These
standard potentials only utilise the temperature dependence of the
pressure $p$ and the expectation value $L$, for a comparison see
e.g.~\cite{Schaefer:2009ui}. More recently, the quadratic fluctuations
of the Polyakov loop have also been computed in \cite{Lo:2013hla}. A
polynomial potential with an additional logarithmic term including the
Haar measure $M_H$ of the $SU(3)$ gauge group describes the lattice
results for the Polyakov loop fluctuations remarkably well. It reads
\begin{align}\nonumber 
  V_\text{glue}(L,\bar L) &= -\frac{a(T)}{2} \bar L L + b(T)
  \ln M_H(L,\bar L)\\[2ex]
  &\quad + \frac{c(T)}{2} (L^3+\bar L^3) + d(T) (\bar L L)^2\,,
\label{eq:polpot}\end{align}
with the Haar measure as a function of the Polyakov loop and its conjugate,
\begin{align}
M_H (L, \bar{L})= 1 -6 \bar L L + 4 (L^3+\bar L^3) - 3  (\bar L L)^2\,.
\end{align}
The temperature-dependent coefficients in \eqref{eq:polpot} can be
expressed by the following parameterisation
\begin{equation}
  \label{eq:9}
  x(T ) = \frac{x_1 + x_2/t + x_3/t^2}{1 + x_4/t + x_5/t^2}\,,
\end{equation}
for $x \in \{ a,c,d\}$ and $t=T/T_0$ whereas $b(T)$ reads
\begin{equation}
  \label{eq:1}
  b(T ) = b_1 t^{-b_4} (1 -e^{b_2/t^{b_3}} )\ .
\end{equation}
For the deconfinement temperature we use $T_0 =250$ MeV.
The coefficients are collected in Tab.~\ref{tab:coeffs}.
\begin{table}[tb!]
  \centering
  \begin{tabular}{c||c|c|c|c|c}
     & 1 & 2 & 3 & 4 & 5 \rule{0pt}{2.6ex}\rule[-1.2ex]{0pt}{0pt}\\ \hline\hline
    $a_i$ &-44.14& 151.4 & -90.0677 &2.77173 &3.56403 \\\hline
    $b_i$ &-0.32665 &-82.9823 &3.0 &5.85559  &\\\hline
    $c_i$ &-50.7961 &114.038 &-89.4596 &3.08718 &6.72812\\\hline
    $d_i$ & 27.0885 &-56.0859 &71.2225 &2.9715 &6.61433\\
  \end{tabular}
  \caption{Coefficients of the Polyakov loop potential
    parameterisation Eqs.~\eqref{eq:9} and \eqref{eq:1}.} 
  \label{tab:coeffs}
\end{table}
Note in this context that in \cite{Fu:2015naa} it was shown
analytically that the higher moments of the baryon number distribution
crucially depend on the Polyakov loop propagators. Furthermore, the
Polyakov loop susceptibilities computed in \cite{Lo:2013hla} are
proportional to the connected two-point functions of the loops. They
are therefore directly related to their propagators. Thus, since this
parametrisation is optimised for the description of the Polyakov loop
propagators, which in turn are the crucial contributions of the pure
glue sector of QCD to the baryon number fluctuations, the potential in
\Eq{eq:polpot} is the most natural choice for the present purpose.

This potential reduces to the polynomial potential for $b(T) \!=\! 0$,
i.e. when the logarithmic term is dropped. While both
parameterisations give the same results for $T \!<\! T_c$, the results
for the Polyakov loop susceptibilities deviate largely for $T \!>\!
T_c$. Hence, based on the discussion above, we expect improvements of
the previous results in Ref.~\cite{Fu:2015naa,Fu:2015amv} on the baryon number
fluctuations for $T > T_c$. We demonstrate this explicitly in
Sec.~\ref{sec:num}.


\section{Correlation functions at
  finite density and Silver Blaze}\label{sec:fre}


\subsection{Silver Blaze Property and the Frequency Dependence
}\label{sec:sbprop}

At vanishing temperature the quark chemical potential has to exceed a
critical value $\mu_c$ before the system can reach a finite
density. This is known as the Silver Blaze property
\cite{Cohen:2003kd}. In the context of the FRG, the consequences of
the Silver Blaze property have been discussed first in
\cite{Khan:2015puu,Fu:2015naa} and we refer to this work for a more
thorough discussion.

As a consequence, all QCD observables at $T=0$ are independent of the
chemical potential for $\mu \leq \mu_c$. The quark critical chemical
potential $3 \mu_c = M_N-\epsilon_b$ is close to the pole mass of the
lowest lying state with non-vanishing baryon number, i.e., the nucleon
mass $M_N$. The subtraction $\epsilon_b$ accounts for the binding
energy of nuclear matter. In the present work we drop the small
binding energy $\epsilon_b$ and also identify $M_N=3 M_q$. Formally,
this property entails for $\mu \leq \mu_c$ that the $\mu$-dependence
of the correlation functions is given by a simple shift of the
frequency arguments. For the present discussion it is convenient to
decompose the scale-dependent 1PI finite density correlation functions
\begin{align}\label{eq:Gn}
  \Gamma_{\Phi_1\cdots \Phi_n,k}^{(n)}(p_1,\dots,p_n;\mu) =
  \frac{\delta^n\Gamma_k}{\delta\Phi_1(p_1)\cdots\delta\Phi_n(p_n)}\,,
\end{align}
into the vertex function and the momentum conservation, 
\begin{align}\nonumber 
  &\Gamma_{\Phi_1\cdots \Phi_n,k}^{(n)}(p_1,\dots,p_n;\mu)\\[2ex] 
 =&\,
  \tilde\Gamma_{\Phi_1\cdots
    \Phi_n,k}^{(n)}(p_1,\dots,p_{n-1};\mu)\,(2\pi)^4 \delta(p_1+\cdots
  p_n)\,.
\label{eq:decompose}\end{align}
The $\delta$-function involves that we count all momenta as incoming.
Then the Silver Blaze property is entailed in simple equations for
the vertex functions \eq{eq:decompose} for $\mu < \mu_c$,
\begin{align}\label{eq:sbprop}
  \tilde\Gamma_{\Phi_1\cdots
    \Phi_{n},k}^{(n)}(p_1,\dots,p_{n-1};\mu)=\tilde\Gamma_{\Phi_1\cdots
    \Phi_{n},k}^{(n)}(\tilde p_1,\dots,\tilde p_{n-1};0)\,,
\end{align}
with the shifted Euclidean four-momenta
\begin{align}\label{eq:tildep}
\tilde p_j = (p_{0j}+i \alpha_j \mu,\vec{p}_j)\,, \quad
  j=1,\ldots,n\,.
\end{align}
The baryon number of the corresponding fields $\Phi_j$ is given by
$\alpha_j/3$. As the baryon number of all $\Gamma_k^{(n)}$ vanishes we
have $\tilde p_1 +\cdots + \tilde p_n=p_1+\cdots +p_n$, the sum of the
$\tilde p_i$ is real. Hence the property \eq{eq:sbprop} formally
extends to the full correlation functions with $\delta(\tilde
p_1+\cdots + \tilde p_n)$ being well-defined.

The intimate relation between the Silver Blaze property and the
frequency dependence of $n$-point functions is manifest in
Eq.~(\ref{eq:sbprop}). This has important consequences for consistent
approximation schemes: if the frequency dependence of correlation
functions is not taken into account properly, the Silver Blaze
property is violated. This applies to all $n$-point functions
involving legs with nonzero baryon number.

In the present work, these are the running Yukawa coupling $h_k$ and
the quark anomalous dimension $\eta_{q,k} = -\frac{\partial_t
  Z_{q,k}}{Z_{q,k}}$. For example, in the spirit of \eq{eq:sbprop} the
Yukawa coupling reads for $\mu\leq \mu_c$
\begin{align}\label{eq:G3yuk}
  \tilde\Gamma^{(3)}_{q \bar q \phi,k}(p_1,p_2;\mu) \propto h_k(\tilde
  p_1,\tilde p_2)\,,
\end{align}
with $\tilde p_i$ as in \eq{eq:tildep} with $\alpha_{1}=-\alpha_2 = 1$
and $\alpha_3=0$, and we have dropped all terms with other tensor
structures of the vertex. This entails that $\tilde p_1 +\tilde p_2 =
p_1 +p_2 $ and $\tilde p_3 =p_3$.  Note again that all momenta $p_i$
are incoming momenta. The $\mu$-dependent frequencies for quark and
anti-quark read $\tilde p_{01} = p_{01}+ i \,\mu$, $\tilde p_{02} =
p_{02}-i\, \mu$. Similarly it follows for the quark anomalous
dimension
\begin{align}\label{eq:G2etaq}
  \partial_t \tilde\Gamma^{(2)}_{q \bar q,k}(p;\mu) \propto \eta_{q,k}(\tilde 
  p)\,\gamma_\mu \tilde p_\mu\,,
\end{align}
using the momentum conservation $\tilde p_1 = -\tilde p_2$. As in
\eq{eq:G3yuk} we have dropped terms with further tensor structures in
\eq{eq:G2etaq}, here it is only the scalar one proportional to the
quark mass and its flow. We conclude that for $\mu<\mu_c$ we have 
\begin{align}\label{eq:mu0}
\left. \partial_\mu\right|_{\tilde p_i} h_k(\tilde
  p_1,\tilde p_2)= \left.\partial_\mu\right|_{\tilde p}
  \eta_{q,k} (\tilde
  p)=0\,.
\end{align}
Hence, neither the wave function renormalisation $Z_q$ nor the Yukawa
coupling receive a genuine $\mu$-dependence. At vanishing temperature
the density or chemical potential contributions to the flow equation
are proportional to the step function $\Theta(\mu-E_{q,k})$, with the
quasi-particle energies for the quarks $E_{q,k} \!=\!  \sqrt{k^2
  \!+\!M_{q,k}^2}$. For $\mu>\mu_c$ the density contributions are
non-vanishing. Furthermore, the explicit $\mu$-dependence cannot be
accounted anymore for by a shift of the momentum arguments as in
\Eq{eq:sbprop}. This results in manifestly $\mu$-dependent correlation
functions.

Of course, for $T>0$ the step function becomes a Fermi distribution
function $n_F(\bar m_{q,k}^2; T,\mu)$ defined in \eq{eq:nbarF}. So
strictly speaking QCD has no Silver Blaze property at finite
temperature. It is nonetheless crucial for the correct
$\mu$-dependence of the theory to carefully evaluate the frequency
dependence of finite temperature correlation functions. For physical
observables, they enter through the corresponding loop diagrams, for
example through the quark loop contribution to the effective
potential. In this case, their frequency dependence has to be taken
into account in the loop integration. This is discussed in detail in
the next section. If one is interested in the $n$-point functions
themselves, they should be evaluated at complex frequencies $p_{0j} -
i\,\alpha_j\, \mu$ in order to retain the Silver Blaze property. This
can be seen explicitly e.g. in \eq{eq:FB11}. Such an evaluation point
guarantees that the correlation functions are defined at
$\mu$-independent points in momentum space. We want to emphasise that,
in any case, frequency-independent approximation schemes will always
lead to a certain Silver Blaze violation and a corresponding
inaccurate dependence on the chemical potential.

  Moreover, the complex frequency argument $p_0 + i \mu$ in
  \eq{eq:FB11} renders $h_k$ and $\eta_{q,k}$
  complex-valued. Through the corresponding loop diagrams these
  quantities appear in thermodynamic observables or particle
  masses which, in turn, have to be real-valued. In the next
  section we demonstrate explicitly at the example of $\eta_{q,k}$
  that in due consideration of the frequency dependence real-valued
  quantities are finally obtained.

\subsection{Improved Effective Potential}
\label{sec:freqpot}

Since higher moments of the baryon number distribution are defined via
chemical potential derivatives of the effective potential, see \Eq{eq:bnf}, it is of
major importance that the $\mu$-dependence of the effective potential
is resolved properly. As mentioned above, this is intrinsically tied
to the Silver Blaze property of QCD and the frequency dependence of
the quark-involved correlation functions.  More specifically, the
correlation functions that drive the flow of the effective potential
are the ones for the Yukawa coupling $h_k$ and for the quark anomalous
dimension $\eta_{q,k}$. The former quantity enters the flow of the
effective potential  through the quark mass $m_{q,k} \!=\!  h_k
\sigma_0/2$ and the latter through the scale derivative of the quark
regulator, $\partial_t R_k^q(\vec{p}\,) \propto \vec{\gamma} \vec{p}\,
(\partial_t - \eta_{q,k}) r_F(\vec{p}\,)$, with the fermionic
regulator shape function $r_F$. Consequently, it is the quark loop
contribution to the potential flow where the frequency dependence
needs to be taken into account thoroughly.

Furthermore, only the spatial three-momenta and not the frequencies
are regularised. The ensuing non-locality in frequency requires the
full frequency dependence of correlation functions for quantitative
precision. In turn, we use $\vec p=0$, which has been shown in
\cite{Helmboldt:2014iya} to be a good approximation.

Now we apply this reasoning to the quark correlation functions in the
present approximation \eq{eq:action}, $Z_{q,k}, h_k$. Note that the
flow of both these couplings can be deduced from that of the
quark--anti-quark two-point function, see
\cite{Pawlowski:2014zaa,Fu:2015naa}. In the present approximation this
two-point function reads at vanishing pion fields, $\vec \pi=0$,
\begin{align}\label{eq:Gqbarq}
  \tilde \Gamma_{q\bar q,k}^{(2)}(p)= Z_{q,k}(p)\,\left(\pslash
    +\frac{\bar h_k(p) \bar \sigma}{2}\right)\,,
\end{align}
where $\tilde\Gamma_{q \bar{q},k}^{(2)}(p)$ is the two point function
without the momentum-conserving $\delta$-function, see
\eq{eq:decompose}. We have also introduced the normalised couplings
\begin{align}\label{eq:barcoup}
  \bar h_k(p)= \0{ h_k(p,-p)}{Z_{q,k}(p)
    Z_{\phi,k}^{1/2}(0)}\,,\qquad  \bar
  \sigma=Z_{\phi,k}^{1/2}(0) \sigma\,, 
\end{align}
in the spirit of the present derivative expansion of the mesonic
sector.  Now we retain the frequency dependence of the dispersion via
a quark wave function renormalisation $Z_{q,k}(p_0)$. As discussed
before, its spatial momentum dependence is well-captured by the
$k$-dependence of $Z_k$, for a detailed study see
\cite{Helmboldt:2014iya}. There it has been shown that $Z_k(p) \approx
Z_k(0)$ is a quantitative approximation for the regularised momentum
directions.  The Yukawa term $\bar h_k(p) \bar \sigma/2$ relates to the
momentum-dependent renormalisation group invariant (but
cutoff-dependent) mass function $M_{q,k}(p)$ of the quark. This
quantity has been studied with the FRG in \cite{Mitter:2014wpa} in
vacuum QCD. From \cite{Mitter:2014wpa,Braun:2014ata} we deduce that
its momentum-dependence for the current cutoff scales of $k\lesssim
700$ MeV is negligible, and we resort to a momentum-independent
approximation.  Naively this suggests $p=0$. However, in order to
guarantee the Silver Blaze property we evaluate the flow of $\bar
h_k(p)$ at ${\rm Im}\, p_0=-i\,\mu$ and $\vec p=0$. The real part of
$p_0$ is adjusted for capturing the correct thermal decay, the details
are given below. This ensures that the dependence on $\tilde p$ is not
confused with a genuine $\mu$-dependence of $\bar h_k$, see the
discussion in the previous chapter and \cite{Fu:2015naa}.

This leaves us with the task of calculating the frequency-dependent
quark anomalous dimension $\eta_{q,k}(p_0)$. It is obtained from the
flow of the quark two-point function \eq{eq:Gqbarq} by the following
projection prescription
\begin{align}
  \eta_{q,k}(p)=&\frac{1}{Z_{q,k}(p)}\frac{1}{4 N_c
    N_f}\frac{\partial^2}{\partial |\vec{p}|^2}\mathrm{Tr}\bigg(i
  \vec{\gamma}\cdot\vec{p}\,\partial_t \tilde\Gamma^{(2)}_{q\bar
    q,k}(p)\bigg)\,.
\label{eq:etapsi}\end{align}
By only keeping the frequency dependence and ignoring the spatial
momenta by setting $\vec{p}=0$, we
arrive at
\begin{align}\nonumber 
  \eta_{q,k}(p_0)=&\frac{1}{24\pi^2N_{f}}(4-\eta_{\phi,k})\bar{h}_{k}^{2}\\[2ex]
  \nonumber &\times\Big\{(N_{f}^{2}-1)
  \mathcal{FB}_{(1,2)}(\bar{m}_{q,k}^{2},\bar{m}_{\pi,k}^{2};p_{0})\\[2ex]
  &
  +\mathcal{FB}_{(1,2)}(\bar{m}_{q,k}^{2},\bar{m}_{\sigma,k}^{2};p_{0})
  \Big\}\,. 
\label{eq:etapsiexp}
\end{align}
The threshold function $\mathcal{FB}_{(1,2)}$ is defined in
App.~\ref{app:threshold}. \Eq{eq:etapsiexp} depends on the meson
anomalous dimension $\eta_{\phi,k}(q)$. It has been evaluated at
vanishing spatial momenta and frequency, $q=0$. In this approximation
it has been derived in \cite{Pawlowski:2014zaa} and reads,
\begin{align}\label{eq:etaphi}
\nonumber
  \eta_{\phi,k}=&\frac{1}{6\pi^2}\Bigl\{\frac{4}{k^2} \bar\kappa_k\,
  (\bar{V}_{k}''(\bar\kappa_k))^2\, \mathcal{BB}_{(2,2)}\left(\bar{m}_{\pi,k}^2,
    \bar{m}_{\sigma,k}^2\right) \Bigr.\\ \nonumber
  &+\Bigl. N_c \bar{h}_{k}(\bar\kappa_k)^2\left[\left(2
      \eta_{q,k}-3\right)\mathcal{F}_{(2)}(\bar{m}_{q,k}^2)\right.\Bigr.\\
  &-\Bigl.\left.4\left(\eta_{q,k}-2\right)
    \mathcal{F}_{(3)}(\bar{m}_{q,k}^2) \right] \Bigr\}\,.
\end{align}
with the threshold function
\begin{align}
  \mathcal{BB}_{(2,2)}\left(\bar{m}_{\pi,k}^2,
    \bar{m}_{\sigma,k}^2\right) = \frac{T}{k} \sum_{p_0}
  G_{\pi\pi}^2(p) G_{\sigma\sigma}^2(p)\,,
\end{align}
and $\mathcal{F}_{(n)}$ given in \Eq{eq:ths1}. The explicit analytic
expressions are given in \cite{Pawlowski:2014zaa,Fu:2015naa}. In
\eq{eq:etaphi}, the effective action is expanded about $\bar\rho =
\bar\kappa_k$ with $\bar\kappa_k=Z_{\phi,k} \kappa$. In contrast to
the standard expansion about the flowing minimum, we use a Taylor
expansion about a fixed $\rho=\kappa$ with $\partial_t \kappa=0$, as
put forward in \cite{Pawlowski:2014zaa}. Hence,
$\partial_t\bar\kappa_k = -\eta_{\phi,k} \bar\kappa_k$.

Note that the quark anomalous dimension $\eta_{q,k}(p_0)$ is in
general complex-valued at finite chemical potential. As discussed in
the previous section, this is related to the fact that correlation
functions that involve quarks are functions of $p_0+i\mu$, which again
is related to the Silver Blaze property. Also, inserting
$\eta_{q,k}(p_0)$ in \eq{eq:etapsiexp} into the flows of other
couplings or the effective potential leads to two-loop frequency
resummations that properly take into account the non-regularised
frequency dependence of the flows.

As already discussed above, for the Yukawa coupling $\bar h_k(p)$ we
follow the procedure in \cite{Fu:2015naa}: We evaluate the coupling at
a fixed external frequency ${\rm Im}\, p_0 + i\mu=0$, and $\vec p=0$.
The real part of $p_0$ is fixed by the requirement that $h_k$ has to
be temperature-independent at energy scales that exceed the thermal
scale, i.e. $k\gtrsim \pi T$ and depends only on the lowest Matsubara
mode for $k\lesssim \pi T$. This suggest $p_0^2 = k^2 + (\pi T)^2
\Theta_T(k/T)$. The comprehensible choice $\Theta_T(x) = \exp(-2x/5)$
distinguishes between the low- and high-energy regime relative to the
thermal scale. This can be viewed as a phenomenologically motivated
procedure to circumvent the necessity of a fully frequency dependent
Yukawa coupling $\bar h_k$.  The flow of $\bar h_k$ also depends on
the frequency-dependent anomalous dimension $\eta_{q,k}(q_0)$ with the
loop frequency $q_0$, leading to two-loop contributions in the flow of
the Yukawa coupling. In the context of the present work the latter is
only relevant for the flow of the effective potential, where the
frequency resummations in the flow of the Yukawa coupling relate to
three-loop frequency effects which we consider to be sub-leading. We
therefore drop the frequency dependence of $\eta_{q,k}$ in the flow of
$\bar h_k$ and use the same approximation as for the frequency
dependence of $\partial_t \bar h_k$ also for the quark anomalous
dimension in the diagram. This leads us to the same flow for $\bar
h_k$ as used in \cite{Fu:2015naa},
\begin{align}\label{eq:flowbarh}
\begin{split}
  \partial_t\bar{h}_k &=\Bigl( \frac{1}{2}\eta_{\phi,k}+ \eta_{q,k}
  \Bigr)\bar{h}_k +\frac{1}{4\pi^2 N_f}\bar{h}_k^3\\
  &\quad\times \left[
    L_{(1,1)}^{(4)}\left(\bar{m}_{q,k}^2,\bar{m}_{\sigma,k}^2,\eta_{q,k},\eta_{\phi,k};p_0\right)\right.\\
  &\quad\left.-(N_f^2-1)\,
    L_{(1,1)}^{(4)}\left(\bar{m}_{q,k}^2,\bar{m}_{\pi,k}^2,\eta_{q,k},\eta_{\phi,k};p_0\right)\right]\,,
\end{split}
\end{align}
with
\begin{align}
\begin{split}
  L_{(1,1)}^{(4)}=\frac{2}{3} \left[\!
    \left(1-\frac{\eta_{\phi,k}}{5}\right)\!\mathcal{FB}_{(1,2)}+\!\left(1
      -\frac{\eta_{q,k}}{4}\right)\!\mathcal{FB}_{(2,1)} \right]\,.
\end{split}
\end{align}
We omitted the arguments for the sake of brevity here. They are the
same as in \eq{eq:etapsiexp} and are defined in
App.~\ref{app:threshold}. Note that the full Yukawa coupling
$h_k(p_0)= Z_{q,k}(p_0) \bar h_k$ carries the relevant frequency
dependence.

\begin{figure}[t]
\includegraphics[scale=0.22]{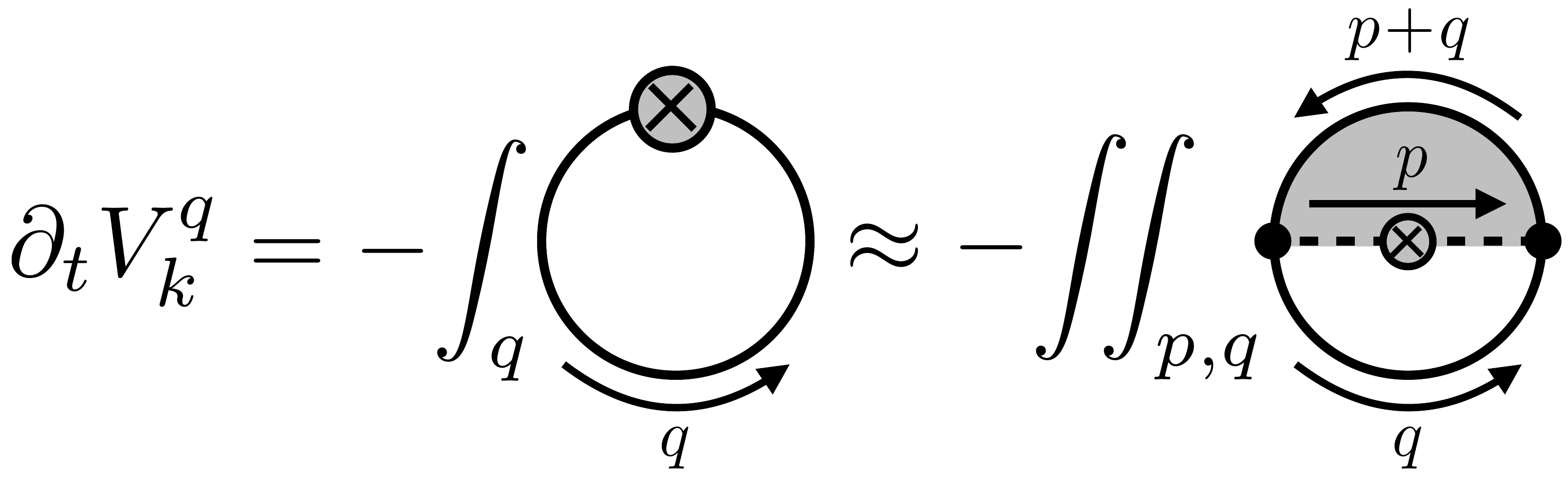}
\caption{Simplified illustration of the quark loop contribution to the flow of
  the effective potential. The first loop represents the standard form
  of the quark contribution. The crossed circle is the regulator
  insertion and the loop momentum integral also includes the frequency
  summation. The last term corresponds to $\Delta \partial_t V_k^q$ in \Eq{eq:vstruc}. It illustrates how the full frequency
  dependence of the quark anomalous dimension enters here. The gray
  area corresponds to the contribution from the quark anomalous
  dimension. We dropped $\partial_t V_k^q\Bigr|_{\eta_{q,k}=0}$ in the last term for the sake of simplicity.
}\label{fig:qloop}
\end{figure}

Finally, we discuss the flow equation of the effective potential. This
is now derived in the presence of the frequency dependence of
$Z_{q,k}(p_0)$.  To illustrate this procedure, we first take a closer
look at the structure of this equation. One can rewrite the flow of
the quark contribution, $\partial_t V_k^q$ to the effective potential (up to a volume factor) as
\begin{align}\label{eq:vstruc}\nonumber
  \partial_t V_k^q=& 
  -\tr\,T\sum_{q_0}\!\int_{\vec{q}} G_{\bar q q}(q_0,\vec{q}\,) \,
  \partial_t R_k^q(q_0,\vec{q}\,) \\[2ex] \nonumber =&
  - \tr\,T\sum_{q_0}\!\int_{\vec{q}} G_{\bar q q}(q_0,\vec{q}\,)\,
  \vec{\gamma} \vec{q}\, \bigl(\partial_t - \eta_{q,k}(q_0)\bigr)
  r_F(\vec{q}\,)\\[2ex]
  =&\partial_t V_k^q\Bigr|_{\eta_{q,k}=0} +\Delta \partial_t V_k^q\,,
\end{align}
where the trace sums over color, flavor and spinor indices. In the
last line of \eq{eq:vstruc} we have split the flow into a contribution
with and without the quark anomalous dimension.  The frequency
summations carry a two-loop structure due to the frequency-dependent
quark anomalous dimension $\eta_{q,k}(p_0)$. The quark propagator that
enters in $\eta_{q,k}$ carries both frequencies via $q_0+p_0$. This
is illustrated in Fig.~\ref{fig:qloop}. In addition, the color trace
has to be performed after the frequency summation in the presence of a
non-vanishing temporal gluon background. This will be discussed at the
end of this section.

%
%

The two-loop summation can be carried out analytically and we still
arrive at an analytic expression for the flow equation of the
effective potential,
\begin{align}\nonumber 
  \partial_{t}V_{k}(\rho)&=\frac{k^{4}}{360\pi^{2}}\bigg\{ 12 (
  5-\eta_{\phi,k})\big[(N_{f}^{2}-
  1)\mathcal{B}_{(1)}(\bar{m}_{\pi,k}^{2})\\[2ex]
  \nonumber &\quad+\mathcal{B}_{(1)}(\bar{m}_{\sigma,k}^{2})\big] - 5
  N_c\Big(48 N_f \mathcal{F}_{(1)}(\bar{m}_{F,k}^{2})\\[2ex]\nonumber
  &\quad + \frac{1}{2\pi^{2}}(-4+\eta_{\phi,k})\bar{h}_k^2\Big[
  \mathcal{FFB}_{(1,1,2)}(\bar{m}_{F,k}^{2}
  , \bar{m}_{\sigma,k}^{2})\\[2ex]
  &\quad+(N_{f}^{2}- 1)\mathcal{FFB}_{(1,1,2)}( \bar{m}_{F,k}^{2} ,
  \bar{m}_{\pi,k}^{2})\Big]\Big)\bigg\}\,,
 \label{eq:Vflow}\end{align}
with the threshold functions, see also \cite{Fu:2015naa}
\begin{equation}
  \label{eq:ths1}
  \mathcal{B}_{(n)} = \frac{T}{k} \sum_{p_0} G_{\phi\phi}^n(p)\quad\text{
    and }\quad   \mathcal{F}_{(n)} = \frac{T}{k} \sum_{p_0} G_{\bar{q}q}^n
  (p)\ .
\end{equation}
The new contributions from the frequency dependent quark anomalous
dimension are the last two lines of \eq{eq:Vflow} with the new
threshold functions $\mathcal{FFB}_{(1,1,2)}$. They encode the
two-loop frequency summation discussed above and read schematically
\begin{align}\label{eq:ffbs}
  \mathcal{FFB}_{(1,1,2)} = \frac{T^2}{k^2} \sum_{p_0}\sum_{q_0}
  G_{\bar q q}(q) G_{\bar q q}(p+q) G_{\phi\phi}^2(p)\,.
\end{align}
This summation can be
performed analytically and the result is given in
App.~\ref{app:threshold}.

It is remarkable that, although $\eta_{q,k}(p_0)$ is complex-valued at
finite $\mu$, the flow equation \eq{eq:Vflow} itself is manifestly
real-valued when the frequency dependence is taken into account. This
is in accordance with our previous discussion: real-valued physical
observables that respect the Silver Blaze property have to comprise
the frequency dependence of the corresponding baryon number carrying
correlation functions.

Finally, we evaluate the color trace which is crucial for the correct
implementation of the Polyakov loop dynamics.  The coupling between
the gauge and the matter sector is achieved by considering a
non-vanishing temporal gluon background field $A_0$. In practice, this
amounts to an imaginary shift of the chemical potential in the
equations for the matter sector, $\mu \rightarrow \mu + i g
A_0$. Hence, one can carry out the Matsubara summation in
\Eq{eq:Vflow} without any reference to the presence of gluons and
simply shift the chemical potential after the summation. However,
since $A_0 = A_0^a t^a$ with $t^a \!\in\! SU(3)$ is in the adjoint
representation, the color trace has to be performed after the
shift. Even though $A_0$ can always be rotated into the Cartan
subalgebra of the gauge group, the color trace is rather involved in
this case due to the two-loop frequency summation. However, it is
always possible to re-express the $A_0$-dependence in favor of the
Polyakov loops $L,\,\bar L$ since the chemical potential enters the
flow equation through Fermi distribution functions $n_F$. The
analytical result of this procedure can be found in
App.~\ref{app:threshold}. Since the glue sector only couples to the
quarks, only the threshold function $\mathcal{FFB}_{(1,1,2)}$ is
involved.

\section{\label{sec:num}Numerical results}

In the last chapter we have derived the flow equation of the effective
potential of the low-energy effective theory, \eq{eq:Vflow} with
\eq{eq:flowbarh} and the quark and meson anomalous dimensions
\eq{eq:etapsiexp} and \eq{eq:etaphi}. It is left to specify the
initial effective action and the UV-cutoff of the effective theory. 

The latter is chosen $\Lambda=700\,\mathrm{MeV}$ in order to keep as
many matter fluctuations as possible while maximising the glue
decoupling on the other hand. See \cite{Fu:2015naa} for more details.
At this initial UV scale we approximate the initial effective 
potential by
\begin{align}\label{eq:VLambda}
  \bar{V}_{\Lambda}(\bar{\rho})=\frac{\bar{\lambda}_{\Lambda}}{2}\bar{\rho}^2
  +\bar{\nu}_{\Lambda}\bar{\rho}\,.
\end{align}
In addition to the two couplings $\bar{\lambda}_{\Lambda}$ and
$\bar{\nu}_{\Lambda}$ the Yukawa coupling $\bar{h}_{\Lambda}$ and the
explicit chiral symmetry breaking parameter $\bar{c}_{\Lambda}$ have to
be provided.

\begin{table}[b]
  \centering
  \begin{tabular}[c]{L|c|c|c|c}
    \hline \hline 
    Truncations &
                  $\bar{\lambda}_{\Lambda}$&$\bar{\nu}_{\Lambda}$[$\mathrm{GeV}^2$] & $\bar{h}_{\Lambda}$ & $ \bar{c}_{\Lambda}$ [$\times 10^{-3}\mathrm{GeV}^3$] \rule{0pt}{2.6ex}\rule[-1.2ex]{0pt}{0pt}\\ \hline
    with frequency dependence &20.7 & 0.24 & 7.2 & 1.96\\\hline
    without & 9.7 & 0.31 & 7.2 & 1.96\\ \hline
  \end{tabular}
  \caption{Input parameters for the truncation with and without 
    frequency-dependent quark anomalous dimension, cf. \Eq{eq:Vflow}.} 
  \label{tab:para}
\end{table}

These initial couplings are determined by fitting the pion decay
constant $f_\pi =92.5\,\mathrm{MeV}$, the pion mass
$m_{\pi}=135\,\mathrm{MeV}$, the $\sigma$-meson curvature mass
$m_{\sigma}=450\,\mathrm{MeV}$, and the quark mass
$m_{q}=297\,\mathrm{MeV}$ in the vacuum.

\begin{figure}[t]
\includegraphics[scale=0.65]{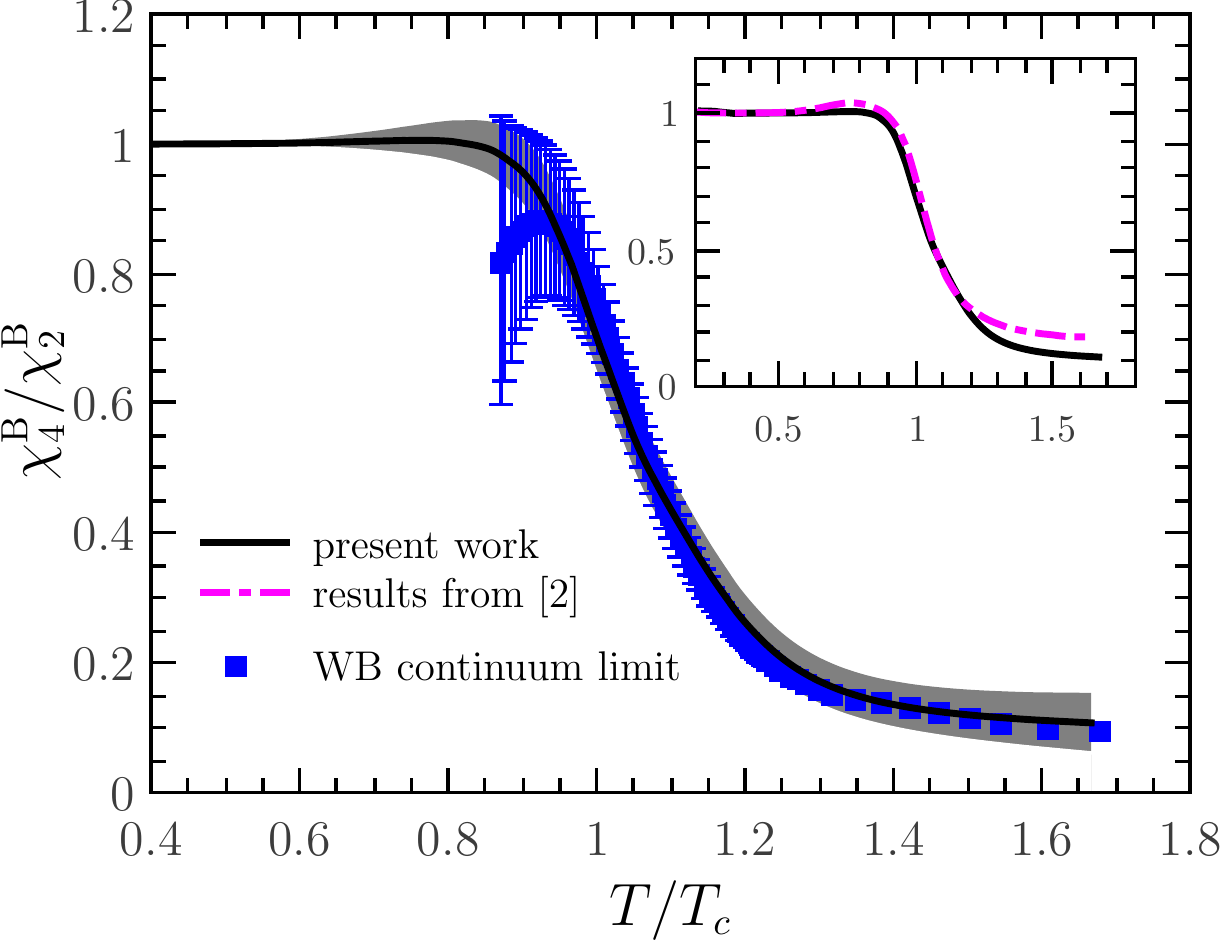}
\caption{Kurtosis
  $\kappa\sigma^2=\chi_4^{\mathrm{B}}/\chi_2^{\mathrm{B}}$ of the
  baryon number distribution in comparison with continuum-extrapolated
  lattice results from the Wuppertal-Budapest collaboration
  \cite{Borsanyi:2013hza}. The gray band shows an error estimate
  according to \Eq{eq:kurtosiserror}. The inlay shows a comparison
  between our present result for the kurtosis and the result of
  \cite{Fu:2015amv}, where neither the frequency dependence, nor the
  Polyakov loop fluctuations have been taken into
  account.}\label{fig:kurtosis}
\end{figure}
%

These observables do not fix the set of initial parameters
completely. This allows us to imprint further QCD information in the
model: as has been discussed in \cite{Fu:2015amv}, the vacuum QCD
flows of the couplings present in the effective theory are known from
\cite{Mitter:2014wpa,Braun:2014ata}. Hence we utilise the remaining
freedom in the set of initial parameters in order to imprint the known
QCD-flow in the large cutoff regime of the effective theory for cutoff
scales $k$ close to $\Lambda$. Effectively this is done by simply
minimising the meson fluctuations for large cutoff scales, as the
mesons quickly decouple at these scales in full QCD. This leads to the
upper parameter set in \tab{tab:para}. For an evaluation of the
fluctuation physics carried by the frequency dependence we compare the
full results with that obtained in the approximation used in
\cite{Fu:2015amv}, see lower parameter set in \tab{tab:para}. Note
that in comparison to \cite{Fu:2015amv} we have also changed the
Polyakov loop effective potential, for the discussion see
chapter~\ref{sec:polloop}.

\Fig{fig:kurtosis} summarises our results for the fluctuations at
vanishing density as a function of temperature: we show the kurtosis
of baryon number distributions, the ratio between the quartic and
quadratic baryon number fluctuations, as a function of the temperature
in comparison with the continuum-extrapolated lattice results from the
Wuppertal-Budapest collaboration \cite{Borsanyi:2013hza}. While our
computation is done in a $N_f=2$ flavor low energy effective theory,
the lattice results are obtained for $N_f=2+1$. Such a comparison
necessitates the introduction of reduced or relative temperatures, and
the absolute temperatures are rescaled by their corresponding
pseudo-critical temperature $T_c$. We have chosen $T_c=155$ MeV for
the $2+1$ flavor lattice simulation, which is obtained from
$\chi_4^{\mathrm{B}}$ simulations in \cite{Bellwied:2015lba}. This
number is also consistent with the calculations in
\cite{Borsanyi:2010bp}. For the $N_f=2$ computations the
pseudo-critical temperature is $T_c = 180$ MeV, which is related to
the maximal magnitude of the derivative of
$\bar{\rho}_{\text{\tiny{EoM}}}$, in the effective potential, with
respect to the temperature.

\begin{figure}[t]
\includegraphics[scale=0.35]{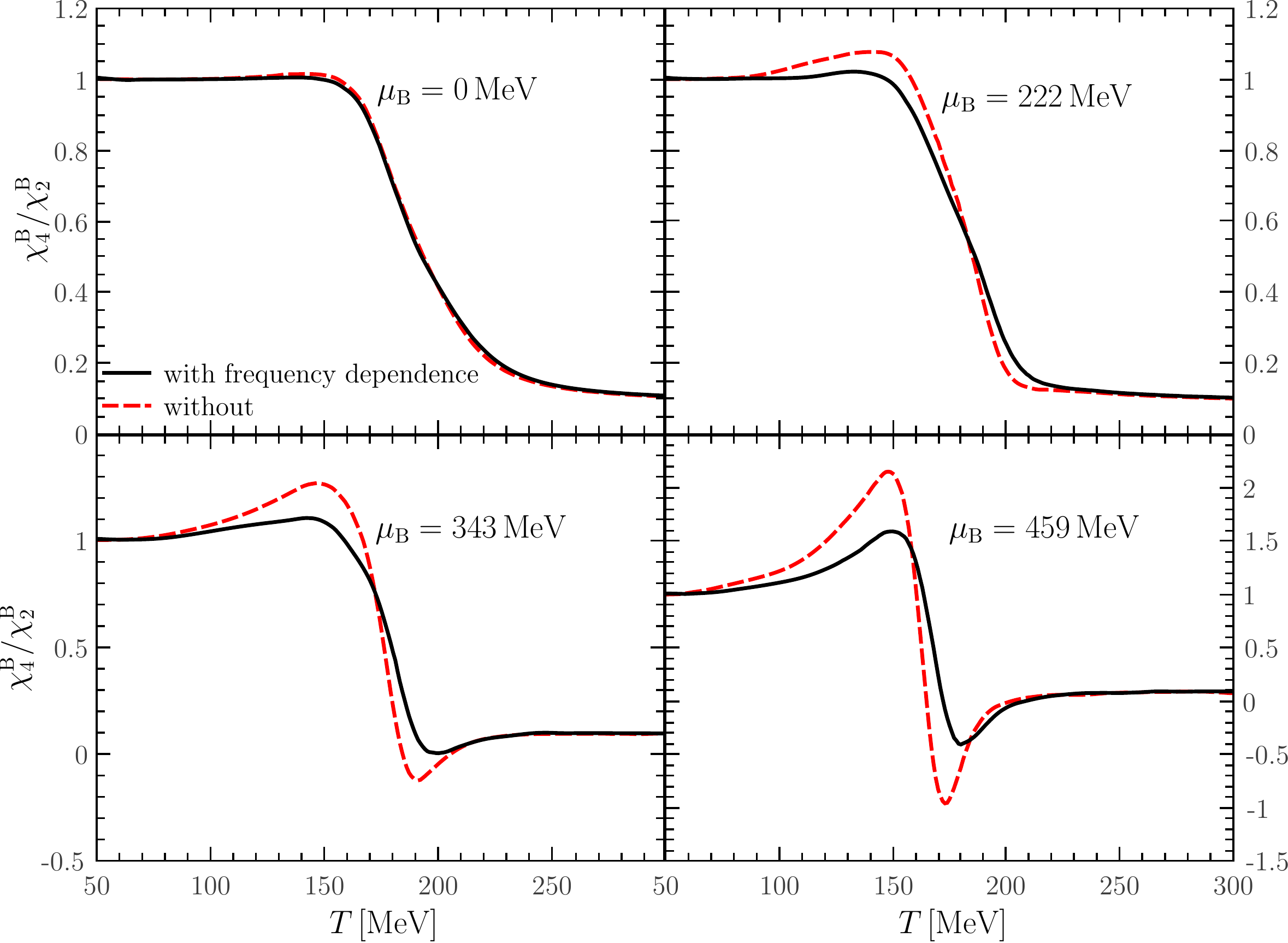}
\caption{Kurtosis as a function of the temperature for different
  baryon chemical potentials. Solid lines include the frequency
  dependence of the quark anomalous dimension and dashed lines not. }
\label{fig:kurtosismuB}
\end{figure}
%
%
\begin{table}[b]
  \centering
  \begin{tabular}[c]{c|c|c|c|c|c|c|c}
    \hline \hline
    $\sqrt{s}\,[\mathrm{GeV}]$ & 200 & 62.4 & 39 & 27 & 19.6 & 11.5 & 7.7\\ \hline
    $\mu_{B,N_f=2}\,[\mathrm{MeV}]$&25.3&78.1&121&168.7&222.7&343&459.4\\ \hline
  \end{tabular}
  \caption{$\mu_{B,N_f=2}$
    corresponding to different collision energy, with \Eq{eq:muBscaleBeta}, for details see \cite{Fu:2015amv}. }
  \label{tab:muB}
\end{table}
The grey band in \Fig{fig:kurtosis} gives a rough estimate of the
systematic error for the computation. It relates to the temperature
dependence of the initial condition of the effective action, and can
be estimated by that of the flow at $k=\Lambda$, for more details see
\cite{Herbst:2013ufa,Fu:2015naa}. This leads to the estimate
\begin{align}\label{eq:kurtosiserror}
  \0{\chi_4^{\mathrm{B}}}{\chi_2^{\mathrm{B}}}\pm \Delta
  \0{\chi_4^{\mathrm{B}}}{\chi_2^{\mathrm{B}}}
  =\0{\chi_4^{\mathrm{B}}}{\chi_2^{\mathrm{B}}}\Big(1\pm
  \0{4}{e^{\Lambda/T}-1}\Big)\,,
\end{align}
with $\Lambda=700\,\mathrm{MeV}$. We find that both the kurtosis
calculated with the frequency dependence and that without, see also
\Fig{fig:kurtosismuB}, agree with the lattice results over the full
temperature range. While the effect of the frequency dependence is
only minor at vanishing density, the Polyakov loop potential we use in
the present work is of major importance at large temperatures. This is
due to the fact that, in contrast to the potential used in
\cite{Fu:2015naa,Fu:2015amv}, this potential also correctly captures
the effect of Polyakov loop fluctuations above $T_c$, see
\cite{Lo:2013hla}. It is precisely this regime where the present
results at vanishing density differ from that in \cite{Fu:2015amv},
see the inlay figure in \Fig{fig:kurtosis}. Moreover, the current
results agree quantitatively with the lattice results. This emphasises
the importance of Polyakov loop fluctuations for the baryon number
fluctuations or more generally higher order correlations as discussed
in \cite{Fu:2015naa}.

In \Fig{fig:kurtosismuB} we compare the dependence of kurtosis on the
temperature at several values of the baryon chemical potential for the
two cases with and without the frequency dependence. Here
$\mu_{\mathrm{B}}=0$, 222, 343, 459 MeV are chosen. We have found that
the difference of the kurtosis irrespective of the frequency
dependence is small at vanishing baryon chemical potential. However,
we argued in Sec.~\ref{sec:fre} that the frequency is intimately
related to the chemical potential dependence and therefore expect that
this will increasingly important with increasing $\mu$. Indeed, we
find that the frequency dependence improved effective potential has a
large effect on the kurtosis at large $\mu$. The frequency dependence
reduces the amplitude of the kurtosis significantly during the
crossover. Our finding implies that the frequency dependence of the
quark anomalous dimension is important and indispensable for STAR, CBM
and HADES-related physics.

Similarly to \cite{Fu:2015amv} we can map our results of the skewness
and curtosis as functions of temperature and chemical potential in
$N_f=2$ flavor QCD to that of the kurtosis at freeze-out temperatures
as a function of the collision energy $\sqrt{s}$ in $N_f=2+1$ flavor
QCD. This is done by an appropriate rescaling of the dimensionful
quantities that captures the different scale-dependence in both
theories. Firstly we adopt the same relation between the chemical
potentials in these theories as \cite{Fu:2015amv}, which is derived
from the experimentally measured skewness
$S\sigma=\chi_3^{\mathrm{B}}/\chi_2^{\mathrm{B}}$ and the
$\sigma^2/M=\chi_2^{\mathrm{B}}/\chi_1^{\mathrm{B}}$ in
\cite{Luo:2015ewa}. This leads to
\begin{equation}\label{eq:muBscaleBeta}
\mu_{B,N_f=2}\approx 1.13\,\mu_{B,N_f=2+1}\,,
\end{equation}
The respective collision energies are summarised in \Tab{tab:muB}. In
this table all collisions energies are far away from the critical
endpoint of the model.

We use the same systematic error estimate as in \cite{Fu:2015amv},
accounting for the uncertainties in determining the freeze-out
temperature, the chemical potential as well as the collision energy in
the present set-up.

\begin{figure}[t]
\includegraphics[scale=0.7]{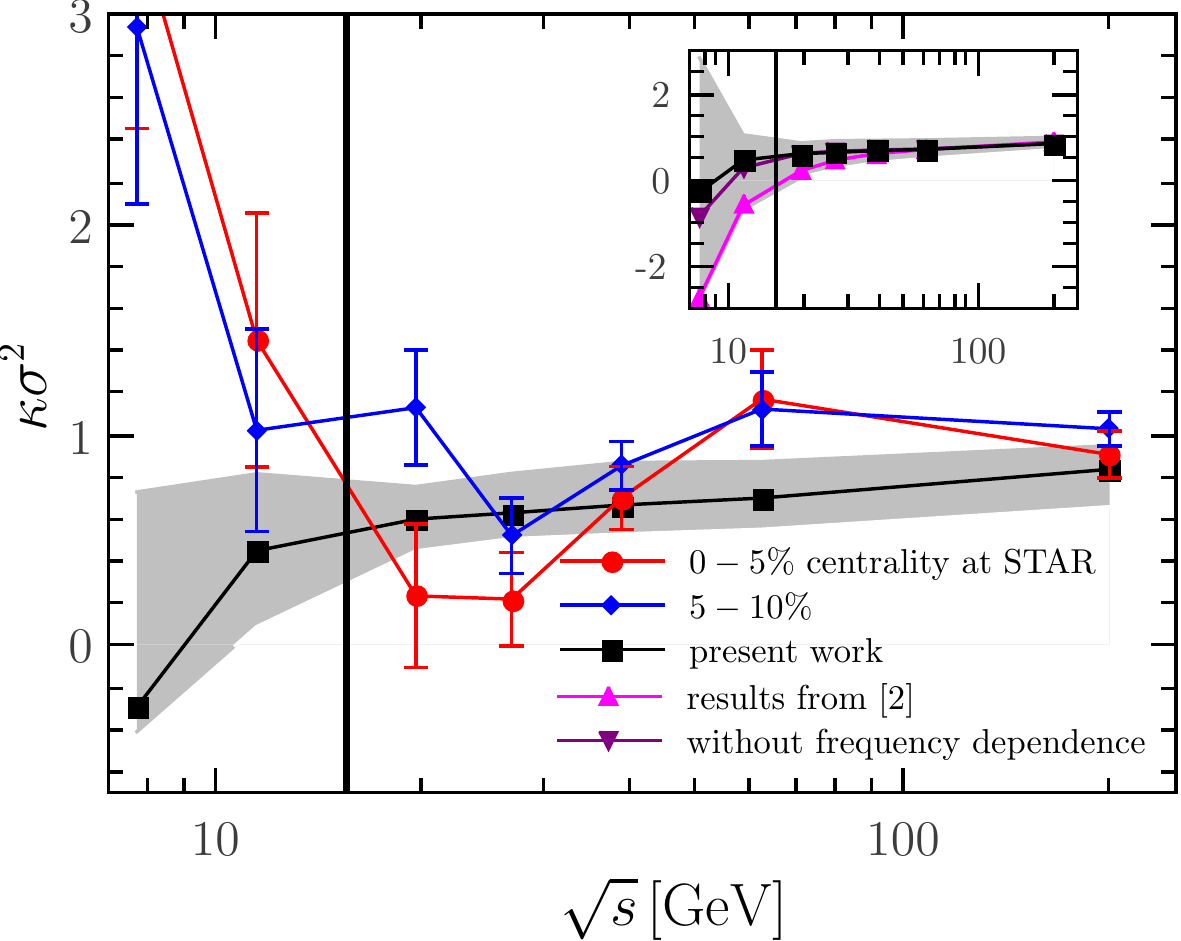}
\caption{Calculated kurtosis $\kappa \sigma^2$ as a function of the
  collision energy, in comparison with experimental measurements in
  $\mathrm{Au}+\mathrm{Au}$ collisions at RHIC with centralities
  $0-5\%$, $5-10\%$ \cite{Luo:2015ewa}. Following \cite{Fu:2015amv},
  we show error estimates resulting from the determination of
  freeze-out temperatures with the gray region, and on the left hand
  side of the black vertical line, the UV-cutoff effect becomes
  significant.}
\label{fig:kurtosissqrts}
\end{figure}

This leads us to \Fig{fig:kurtosissqrts}, the comparison to the
results in \cite{Fu:2015amv} is shown in the inlay. Both results agree
within the respective systematic error bands for collision energies
$\sqrt{s}\gtrsim 19$ GeV. This is the region which has been singled
out by an evaluation of the systematic error in \cite{Fu:2015amv} as
the trustworthy one, and our current results confirm non-trivially
this analysis. For smaller collision energies $\sqrt{s}\lesssim 19$
GeV the systematic errors dominate the result, in
\Fig{fig:kurtosissqrts} the error band only shows the error arising
from the inaccurate determination of the different temperature and
chemical potential scales. 

However, it is the qualitative improvement of the current set-up
  in comparison to \cite{Fu:2015amv}, that already allows for
  interesting conclusions: while in the earlier work the experimental
  results were compatible with the computation also for
  $\sqrt{s}\lesssim 19$ GeV due to the large error bands, the current
  findings clearly deviate at these collision energies. Possible
  important sources of the incompatibilities in this regime are, on
  the one hand, omitted effects, such as the missing high density
  off-shell degrees of freedom in our present
  computations. Potentially, they have an significant impact on the
  existence and location of a possible critical endpoint, as well as
  on the size of the critical region. Furthermore, the lack of
  non-equilibrium effects, see e.g.\ \cite{Herold:2016uvv}, has to be
  remedied. On the other hand, the centrality dependence, or, more
  accurately, the dependence on the $p_T$-cut of the experimental
  results has to be taken into account, see e.g.\
  \cite{Bzdak:2016qdc,Bzdak:2016sxg}. Generally speaking, there are
  further non-critical sources of fluctuations that affect the
  measured baryon number multiplicity distributions and are not
  completely accounted for in the data.  For a recent summary of these
  issues we refer to \cite{Nahrgang:2016ayr} and references
  therein. The discussion of these effects is deferred to future
  work.

%
%

\section{\label{sec:sum} Summary and conclusions}

In this work we have studied the QCD thermodynamics, the baryon number
fluctuations, and the kurtosis of the baryon number distributions in a
low-energy effective model with fluctuations. Quantum, thermal, and
density fluctuations are included within the framework of the
functional renormalisation group, see \cite{Fu:2015naa,Fu:2015amv}. In
comparison to these previous calculations, qualitative improvements
have been included here. Firstly, we have considered the effects of a
non-trivial quark-dispersion via the inclusion of the frequency
dependence of the quark anomalous dimension. The frequency dependence
has been considered on the level of an analytic resummation in terms
of a cutoff-dependent two-loop Matsubara sum. Secondly, we have used a
Polyakov loop potential that takes care of the second order
correlations of the Polyakov loop, see \cite{Lo:2013hla}.

These qualitative improvements reduce significantly the systematic
error of the current set-up in particular in the regime relevant for
STAR, CBM and HADES measurements. On the more technical side, the
frequency dependence considered here implements naturally the Silver
Blaze property of the theory: at vanishing temperature all correlation
functions show no explicit $\mu$-dependence, the only $\mu$-dependence
is that of the frequency arguments for modes with non-vanishing baryon
number, i.e.  $p_0+ i\mu$ for quarks. We believe, that the technical
set-up put forward in the present work takes into account the relevant
frequency effects on a semi-quantitative level. 

The baryon number fluctuations obtained with
frequency dependence are compared with those without the dependence.
This inevitably also includes non-trivial interactions between quarks
and gluons at higher order, that have been derived here for the first
time. We find the difference between them is mild at vanishing baryon
chemical potential, but increases with the chemical potential, which
implies that the frequency dependence of the quark anomalous dimension
plays an important role in CEP-related physics. Furthermore, our
calculated kurtosis of the baryon number distribution is compared with
the lattice results. Our results are in very good agreement with the
lattice simulations over the full temperature range available.

The above improvements allowed for an update of the comparison of
(equilibrium) kurtosis as a function of the collision energy with the
STAR data, see \Fig{fig:kurtosissqrts}. In comparison to the previous
works the qualitatively reduced error band allows for a more
conclusive analysis in particular at low collision energies (high
density), for the detailed discussion see the discussion below
\Eq{eq:muBscaleBeta}: the situations hints strongly towards missing
effects of the $p_T$-cuts as well as that of non-equilibrium
fluctuations, as well as suggesting a direct study in $N_f=2+1$
flavor QCD. The latter extension is work in progress, and we also
plan to extend the current work towards non-equilibrium effects as
well as an analysis of the $p_T$-cut.

\begin{acknowledgments}

  We thank M.~Mitter and Nils Strodthoff for discussions and work on
  related subjects. This work is supported by the AvH foundation,
  EMMI, the BMBF grants 05P12VHCTG and 05P15VHFC1, the grant
  ERC-AdG-290623, the FWF grant P24780-N27, HIC for FAIR, and the DFG
  via SFB 1225 (ISOQUANT).
\end{acknowledgments}

\appendix

\section{Threshold functions}
\label{app:threshold}

The mixed fermion-boson threshold functions at finite temperature $T$ and
chemical potential $\mu$ are in general defined by a product of $i$ fermion
propagators
\begin{equation}
  \label{eq:2}
  G_F(q, \bar{m}^2_{F,k}) = \frac{1}{(\tilde{q}_0 + i\tilde{\mu})^2 +
    1 + \bar{m}^2_{F,k}}\,,
\end{equation}
and $j$ boson propagators
\begin{equation}
  \label{eq:3}
 G_B(q, \bar{m}^2_{B,k}) = \frac{1}{\tilde{q}^{2}_0 +
    1 + \bar{m}^2_{B,k}}\,,
\end{equation}
with the dimensionless $\tilde{\mu}=\mu/k$, $\tilde{q}_0 = q_0/k$ and
momenta $q_0 = (2 n +1) \pi T$ for fermions and
$q_0 = 2 n \pi T$ for bosons $(n \in \mathbb{Z})$, as
\begin{align}
  \label{eq:4}
  \mathcal{FB}_{(i,j)} &(\bar{m}_{F,k}^2, \bar{m}^2_{B,k};p_0)\equiv
  \nonumber\\[2ex]
  & \frac{T}{k} \sum_n G_F^i (q, \bar{m}^2_{F,k}) G_B^j (p-q ,
  \bar{m}^2_{B,k})\ .
\end{align}
Note that with the $3d$ flat cutoffs used in the present work, the
propagators only depend on the frequencies. For $(i,j) \neq (1,1)$
these threshold functions can be generated by corresponding mass
derivatives from $\mathcal{FB}_{(1,1)}$ as follows
\begin{align}
 \nonumber 
  \mathcal{FB}_{(i,j)} &(\bar{m}_{F,k}^2, \bar{m}^2_{B,k};p_0)=
  \frac{(-1)^{i+j-2}}{(i-1)! (j-1)!} \\[2ex]
  &\frac{\partial^{i-1} }{\partial \bar{m}^{2(i-1)}_{F,k} }
  \frac{\partial^{(j-1)} }{\partial \bar{m}^{2(j-1)}_{B,k} }
  \mathcal{FB}_{(1,1)} (\bar{m}_{F,k}^2, \bar{m}^2_{B,k};p_0)\,.
 \label{eq:5}
\end{align}
The Matsubara sum in the first threshold function
$\mathcal{FB}_{(1,1)}$ can be evaluated analytically
\begin{align}\nonumber 
  \mathcal{FB}_{(1,1)} (\bar{m}_{F,k}^{2},&\bar{m}_{B,k}^{2} \; ;\;
                         p_0 ) = \\[2ex]
  \frac{k^3}{2} \Bigg\{ &-
  \frac{n_B ( \bar{m}_{B,k}^{2};T)} { E_B 
    \Big[  ( ip_0-\mu + E_B )^2 - E_F^2  \Big] } \nonumber\\[2ex]
\nonumber & - \frac{ n_B ( \bar{m}_{B,k}^{2};T ) +1} { E_B \Big[ \Big( ip_0-\mu
    -E_B  \Big)^2- E_F^2  \Big] }\\[2ex]
\nonumber 
  &+\frac{ \bar{n}_F ( \bar{m}_{F,k}^{2};T, \mu) } { E_F \Big[ ( ip_0-\mu- E_F )^2
    -E_B^2  \Big] }\\[2ex]
  &+\frac{n_{F}(\bar{m}_{F,k}^{2};T,\mu)-1}{ E_F  \Big[ ( ip_0-\mu+ E_F )^2 - E_B^2
                           \Big] } \Bigg\} \,.
\label{eq:FB11} \end{align}
In this expression the quasi-particle energies 
\begin{align}
  \label{eq:6}
  E_i = k\sqrt{1 + \bar{m}_{i,k}^2}, \qquad i=F,B\,,
\end{align}
appear with the corresponding distribution functions 
\begin{align}
  \nonumber n_{B}(\bar{m}_{B,k}^{2};T)&=\frac{1}{e^{E_B/T}-1}\ ,
  \\[2ex] \nonumber
  n_{F}(\bar{m}_{F,k}^{2};T,\mu)&= \frac{1}{e^{ (E_F-\mu)/T} +1 }\ ,\\[2ex]
  \bar{n}_{F}(\bar{m}_{F,k}^{2};T, \mu)&= n_{F}(\bar{m}_{F,k}^{2};T,
  -\mu)\,,
\label{eq:nbarF}
\end{align}
In the same way, the threshold function consisting of one boson and
two fermion propagators with different frequencies are defined as 
\begin{align}
  \nonumber 
 & \mathcal{FFB}_{(i,j,k)} (\bar{m}_{F,k}^2, \bar{m}^2_{B,k})\equiv \nonumber\\[2ex]
 & \frac{T^2}{k^2} \sum_{n_p} \sum_{n_{q}} G_F^i (p, \bar{m}^2_{F,k})
  G_F^j (q, \bar{m}^2_{F,k}) G_B^k (p-q , \bar{m}^2_{B,k})\ .
\label{eq:4FFB}\end{align}
Note that there are two separate Matsubara summations involved, which
are indispensable to recover the correct $\mu$-dependency. We have
chosen a slightly different momentum rooting here as compared to
Fig.~\ref{fig:qloop} and \Eq{eq:ffbs}. The result is, of course,
independent of this choice of the rooting as long as momentum
conservation is maintained.

Similar to \eqref{eq:5} higher threshold functions with $(i,j,k) \neq
(1,1,1)$ can be generated by corresponding mass derivatives from
$\mathcal{FFB}_{(1,1,1)}$ like
\begin{align}
\nonumber 
 &\mathcal{FFB}_{(1,1,2)}(\bar{m}_{F,k}^{2},
  \bar{m}_{B,k}^2 )\\[2ex]
  =&-\frac{\partial}{\partial \bar{m}_{B,k}^{2}}
  \mathcal{FFB}_{(1,1,1)}(\bar{m}_{F,k}^{2},\bar{m}_{B,k}^2 )\,
  ,\label{eq:FBc21to22}
\end{align}
which is needed in \Eq{eq:Vflow}.

The first threshold function $\mathcal{FFB}_{(1,1,1)}$ can again be
performed analytically with the result
\begin{align}\nonumber 
  &\mathcal{FFB}_{(1,1,1)}(\bar{m}_{F,k}^{2}, \bar{m}_{B,k}^{2} ) =
  \\[2ex]
  \nonumber&f_{cp1}\Big[(1-\bar{n}_{F})\bar{n}_{(F+B)}+(1-n_{F})(-2+n_{(F+B)})\\[2ex]
  \nonumber&+n_B(-2+\bar{n}_{F}+\bar{n}_{(F+B)}+n_{F}+n_{(F+B)})\Big]\\[2ex]\nonumber
  &+ f_{cp2} \Big[
  \bar{n}_{F}^2+(1-n_{F})^2\Big]+\bigg\{f_{cn1}\Big[-\bar{n}_{F}(1+\bar{n}_{(F-B)})\\[2ex]
  \nonumber
  &-n_{F}(-1+n_{(F-B)})-n_B(-2+\bar{n}_{F}+\bar{n}_{(F-B)}\\[2ex]\nonumber
  &+n_{F}+n_{(F-B)})\Big] + f_{cn2}\bar{n}_F (1-n_{F})\bigg\}\\[2ex]
  &\Big/(3-\bar{m}_{B,k}^{2}+4\bar{m}_{F,k}^{2})\,,\label{eq:F2B1c}
\end{align}
Note that besides the standard distribution functions $n_{B}$, $n_{F}$,
$\bar{n}_{F}$, Eqs.~\eq{eq:nbarF}, several new
functions, which incorporate the nontrivial $\mu$-dependence, emerge 
in $\mathcal{FFB}_{(1,1,1)}$. These new
distribution functions are given explicitly by
\begin{align}
n_{(F\pm B)}&= \frac{1}{e^{(E_F \pm  E_B-\mu)/T} +1} \
           . \label{eq:nFmB}
\end{align}
The other $\bar{n}$ combinations are obtained by the corresponding $n$
functions with the replacement $\mu \to -\mu$.  In \eq{eq:F2B1c}, the
coefficients are defined by
\begin{align}\nonumber 
  f_{cp1}&=\frac{k^4}{ 4 E_B E_F  ( E_B^2 +2 E_B E_F  ) },\\[2ex]
\nonumber 
  f_{cp2}&=\frac{k^4}{4 E_B^2 E_F^2}\ ,\\[2ex]
\nonumber 
  f_{cn1}&= \frac{k^2 (  E_B+2 E_F  )}{ 4
                                                     E_B^2 E_F}  ,\\[2ex]
  f_{cn2}&= \frac{k^2}{2 E_F^2} \,.
\end{align}
It is left to specify the relevant threshold functions, when the
coupling between the gluonic background fields and the quarks is taken
into account. In this case, $\mathcal{FFB}_{(1,1,1)}$ in \eq{eq:F2B1c}
takes the following modified form:
\begin{align}\nonumber 
  &\mathcal{FFB}_{(1,1,1)}(\bar{m}_{F,k}^{2}, \bar{m}_{B,k}^{2} )
  \Big|_{\mathrm{modified}} = \\[2ex] \nonumber
  &f_{cp1}\Big[-\bar{n}_{F,(F+B)}-n_{F,(F+B)}+n_{F}+\bar{n}_{(F+B)}-1\\[2ex]
  \nonumber&+n_B(-2+\bar{n}_{F}+\bar{n}_{(F+B)}+n_{F}+n_{(F+B)})
  \Big]\\[2ex]\nonumber
  &+f_{cp2}(\bar{n}_{F,F}+n_{F,F})+\bigg\{-f_{cn1}\Big[n_{F,(F-B)}+
  \bar{n}_{F,(F-B)}\\[2ex]\nonumber
  &+\bar{n}_{F}+n_{(F-B)}-1+n_B(-2+\bar{n}_{F}+\bar{n}_{(F-B)}\\[2ex]\nonumber
  &+n_{F}+n_{(F-B)})\Big]+f_{cn2}n^{\prime}_{F,F}\bigg\}\\[2ex]
  &\Big/(3-\bar{m}_{B,k}^{2}+4\bar{m}_{F,k}^{2})\,.\label{eq:F2B1cGluon}
\end{align}
In comparison with \eq{eq:F2B1c} there are several new distribution
functions, whose explicit expressions are given by
\begin{align}
  n_{F,(F\pm B)}  &=  \frac{1}{2} \left[ 4 \, n_{2F} \, n_{2(F \pm B)}
    \left( L^2-\bar{L} \right) + n_{1F} \, n_{1(F \pm B)}
  \right. \nonumber\\[2ex]&\quad\times \left(\bar{L}^2-L\right)+(L
  \bar{L}-1) (n_{1F}\,n_{2(F \pm B)}
\nonumber  \\[2ex]
  & \left. \quad + n_{2F} \,n_{1(F\pm B)})+2
    (n_{F}-1)(n_{(F \pm B)}-1)\right] , 
\end{align}
and
\begin{align}
  \bar{n}_{F,(F \pm B)}  &= \frac{1}{2} \left[  4\, \bar{n}_{2F}
    \,\bar{n}_{2(F \pm B)} \left( \bar{L}^2- L\right) +
    \bar{n}_{1F}\,\bar{n}_{1(F \pm B)}
  \right. \nonumber\\[2ex]
&\quad\times \left( L^2-\bar{L} \right) + (
  L \bar{L}-1 ) ( \bar{n}_{1F} \, \bar{n}_{2(F \pm B)} \nonumber
  \\[2ex]
& \left. \quad +\, \bar{n}_{2F} \, \bar{n}_{1(F \pm B)})+2
    \bar{n}_{F}\, \bar{n}_{(F \pm B)}\right] \, ,
\end{align}
\begin{align}
  n_{F,F}&=2 n_{2F}^2
  \left(L^2-\bar{L}\right)+\frac{1}{2}n_{1F}^2\left(\bar{L}^2-
    L\right)\nonumber\\[2ex]&\quad+n_{1F}\,n_{2F}(L
  \bar{L}-1)+(n_{F}-1)^2,\\[2ex]
  n^{\prime}_{F,F}&=\bar{n}_{1F}\,n_{2F}\left(L^2-\bar{L}\right)+
  n_{1F}\,\bar{n}_{2F} \left(\bar{L}^2-L\right)\nonumber\\[2ex]
  &\quad+\frac{1}{4} (L \bar{L}-1)(\bar{n}_{1F}n_{1F}+4\,
  \bar{n}_{2F}\,n_{2F})\nonumber\\[2ex]&\quad-\bar{n}_{F}(n_{F}-1),\\[2ex]
  \bar{n}_{F,F}&=2
  \bar{n}_{2F}^2\left(\bar{L}^2-L\right)+\frac{1}{2}\bar{n}_{1F}^2
  \left(L^2-\bar{L}\right)\nonumber\\[2ex]&\quad+\bar{n}_{1F}\,\bar{n}_{2F}
  (L\bar{L}-1)+\bar{n}_{F}^2\,,
\end{align}
where we have defined
\begin{align}
  \begin{split}
    n_{1F}(x,T,L,\bar{L})&=\frac{2\,e^{x/T}}{\Big(1+3\bar
      L\,e^{x/T}+3L\,
      e^{2x/T}+e^{3x/T}\Big)},\\[2ex]
    n_{2F}(x,T,L,\bar{L})&=\frac{e^{x/T}}{2}\ n_{1F}(x,T,L,\bar{L})\,,
\end{split}
\end{align}
with $i=1,2$
\begin{align}\nonumber 
n_{iF} &=n_{iF}(E_F-\mu,T,L,\bar{L})\ ,\\[2ex] 
n_{i(F \pm B)} &=n_{iF}\left(E_F \pm E_B-\mu,T,L,\bar{L}\right)\,.
\end{align}
The other $\bar{n}$ functions are obtained by replacing
$\mu \to -\mu$, $\bar{L} \to L$ and $L \to \bar{L}$.


\bibliography{ms.bib}

\end{document}